# Autocata: Reaction-network-guided autonomous catalyst screening via capability-centric AI

Ruili Li[1, 5], Rui Qi[2, 3*], Qingqing Mao[4], Ritankar Das[4], Beien Zhu[2, 3*], and Yi Gao[2, 3*]

[1]Key Laboratory of Interfacial Physics and Technology, Shanghai Institute of Applied Physics, Chinese Academy of Sciences, Shanghai 201800, China

[2]Photon Science Research Center for Carbon Dioxide, Shanghai Advanced Research Institute, Chinese Academy of Sciences, Shanghai 201210, China

[3]State Key Laboratory of Low Carbon Catalysis and Carbon Dioxide Utilization, Shanghai Advanced Research Institute, Chinese Academy of Sciences, Shanghai 201210, China

[4]Titan Holdings, Hayward, CA 94541, United States

[5]University of Chinese Academy of Sciences, Beijing 100049, China

E-mail: qir@sari.ac.cn; zhube@sari.ac.cn; gaoyi@sari.ac.cn

## Abstract

Artificial intelligence (AI) has expanded the accessible space for catalyst discovery, yet executing complex screening campaigns still requires constant human intervention to bridge high-level reaction objectives with underlying computational pipelines. Here, we present Autocata, a knowledge-embedded autonomous agent that scales catalytic expertise by encoding multistep reaction pathways into reusable, language-invocable capabilities. Powered by a Transformer pretrained on two million catalyst-adsorbate structures, the system incorporates a capability registry of 79 adsorbate-specialized generative models spanning complex reaction networks across carbon, nitrogen, oxygen, and hydrogen intermediates. Given a high-level natural-language prompt, Autocata deconstructs reaction networks, queries and configures matching generative models, generates candidate surface configurations, enforces physical and geometric validity filtering, and coordinates machine-learning potentials evaluations to rank promising materials. Crucially, the agent adaptively reconfigures workflows when encountering capability boundaries or task bottlenecks. We demonstrate these capabilities across $CH_4$ activation, nitrogen reduction reaction ($N_2$RR), and $CO_2$-to-methanol pathways. By shifting from static toolkits to an adaptive, language-driven computational infrastructure, Autocata lowers the barrier for expert-level catalyst discovery and offers a scalable paradigm for self-driving catalytic research.

## Keywords

Autonomous scientific agents; Heterogeneous catalyst discovery; Reaction networks; Generative models; Computational screening workflows

## Introduction

Discovery of heterogeneous catalysts requires navigating a vast chemical space governed by elemental composition, surface facets, coordination environments, and reaction intermediates[1-6]. Conventional computational approaches, leveraging density functional theory (DFT), machine-learning potentials (MLPs), and high-throughput workflows, have substantially expanded the range of candidate structures that can be evaluate[7-11]. However, these conventional screening methods remain inherently constrained in the chemical and structural space they can explore, as they typically rely on predefined candidate libraries or rule-based surface construction, leaving extensive regions of viable catalytic space unsampled. The challenge becomes even more acute for multi-step reactions, where candidate materials must be evaluated consistently across complex networks of multiple intermediates and competing pathways.

Generative artificial intelligence offers a powerful avenue to overcome these exploratory boundaries by enabling direct sampling from learned atomic and chemical distributions[12-18]. In particular, we recently developed adsorbate-specialized generative models that enable direct sampling of surface-adsorbate structures, generating tens of millions of candidate configurations for target intermediates[19]. Building upon this foundation, our subsequent work extended generative screening to a network-level regime encompassing four key adsorbate intermediates in nitrogen reduction[20]. Nevertheless, specialized generative models covering a broad range of adsorbates across diverse reaction classes remain scarce, severely limiting the applicability of network-level screening across broader catalytic space. Furthermore, current generative workflows present a critical operational bottleneck: translating a catalytic objective into an executable screening pipeline, including deploying specialized models, setting parameters, and adaptively analyzing intermediate workflow states, remains heavily dependent on continuous manual operation and analysis by domain experts. The central bottleneck in AI-assisted catalyst discovery thus shifts from constructing individual generative models to organizing, registering, and orchestrating specialized model capabilities into autonomous, end-to-end discovery campaigns.

Large language model (LLM) agents offer a potential route to promising this gap by connecting high-level natural-language objectives with specialized computational tools through iterative planning and environment-facing actions[21-26]. Chemistry-focused systems such as ChemCrow[25] and Coscientist[26] have extended these capabilities to chemical reasoning, laboratory planning, and automated experimentation. More recently, catalyst-specific agents have begun to emerge: Adsorb-Agent[27] employs LLM-guided reasoning to identify stable adsorption configurations, whereas MASTER[28] integrates hierarchical multi-agent reasoning with atomistic simulation workflows for autonomous heterogeneous catalyst exploration. However, natural-language access alone does not eliminate dependence on scientific expertise. For an agent to conduct catalyst screening rather than merely invoke individual tools, domain knowledge must be encoded into the system to map reaction objectives onto chemically meaningful intermediates, organize available models within an explicit capability registry, represent computational procedures as executable workflow skills, and adapt decisions to intermediate outcomes and capability limitations. The key challenge is therefore not simply to connect language models with scientific tools, but to integrate domain-specific model infrastructure with knowledge-guided autonomous orchestration.

Here, we present Autocata, a knowledge-embedded autonomous agent that converts recurring expertise in computational catalyst screening into reusable capabilities accessible through natural language. Built upon a Transformer pretrained on approximately two million catalyst-adsorbate structures, we systematically constructed an adsorbate-resolved capability registry comprising 79 independently fine-tuned generative models, each tailored to distinct C/H/O/N-containing intermediates across diverse catalytic reaction networks. Autocata integrates this registry with domain rules and executable workflow skills to translate high-level research objectives into adaptive catalyst-screening campaigns, including reaction-network decomposition, model selection, structure generation, validity filtering, and MLP-based candidate prioritization. Crucially, the system adaptively reformulates its search strategies when encountering workflow bottlenecks or coverage limitations in the capability registry. We demonstrate these capabilities across increasing levels of task complexity, from open-ended catalyst generation and constraint-driven adsorbate exploration to multi-intermediate screening for $N_2$RR and capability-aware reformulation of a downstream $CO_2$-to-methanol task. Together, these results establish a capability-centric framework for expert-free, natural-language-driven catalyst screening with adaptive execution across complex reaction networks.

# Results

### 1. Autonomous Workflow for Expert-Free Catalyst Discovery

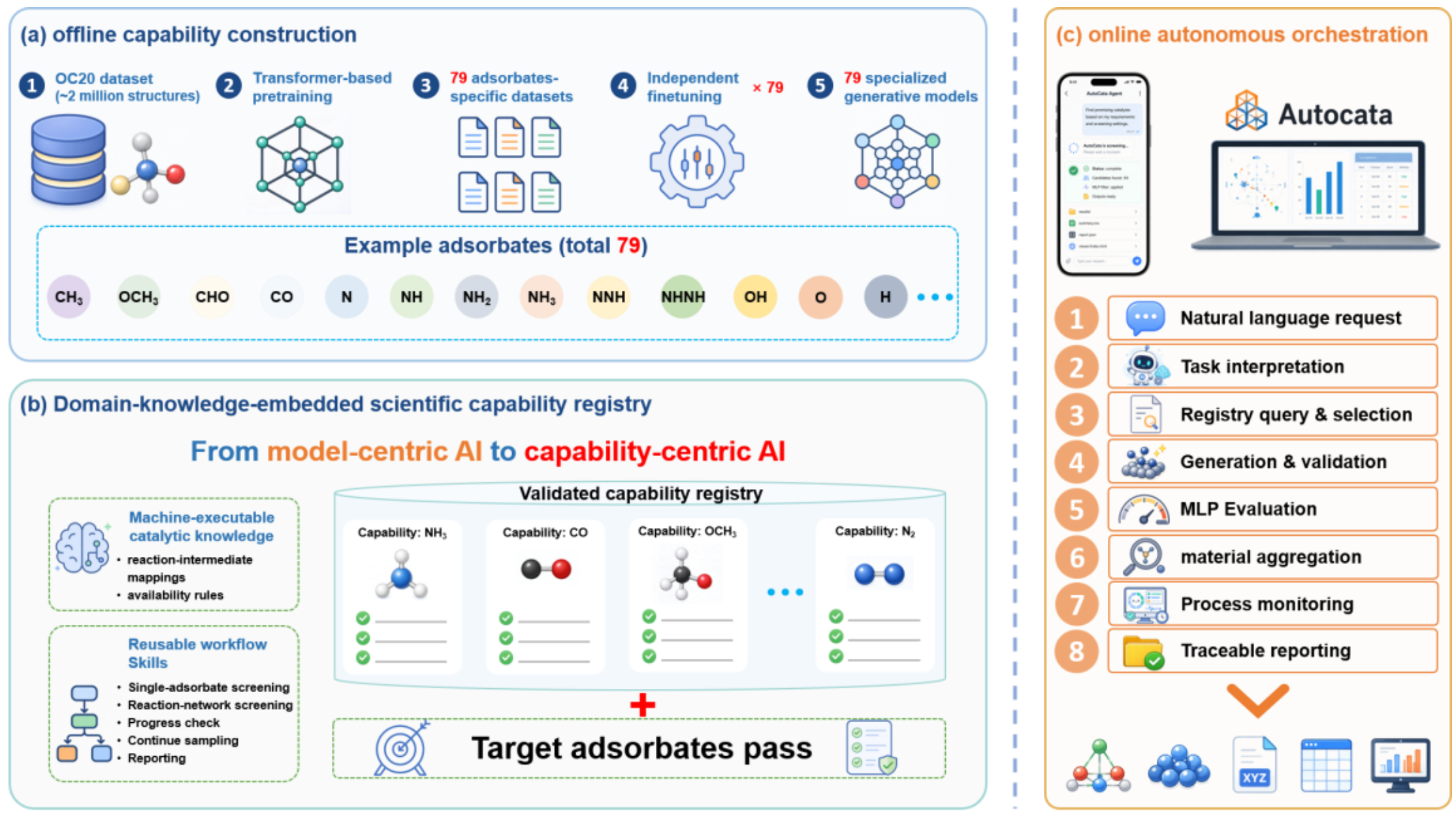

**Figure 1. Capability-centric architecture of Autocata for autonomous catalyst screening. (a)** Offline construction of adsorbate-resolved generative capabilities. **(b)** Domain-knowledge-driven transformation from model-centric AI to capability-centric AI. **(c)** Online autonomous orchestration of catalyst discovery workflows.

Autocata introduces an offline-online architecture that transforms pretrained generative models into reusable scientific capabilities for autonomous catalyst discovery (Fig. 1). In the offline stage, a Transformer-based generative foundation model pretrained on approximately two million catalyst-adsorbate structures from the OC20[29] dataset was systematically adapted into an adsorbate-resolved capability registry. Specifically, we independently fine-tuned the pretrained model for 79 adsorbates spanning carbon-, nitrogen-, oxygen-, and hydrogen-containing intermediates relevant to diverse catalytic reaction networks, generating 79 specialized models with distinct adsorbate-generation capabilities. Their functionality and adsorbate specificity were evaluated through standardized generation tests involving 10,000 structures per model, corresponding to approximately 790,000 generated structures across the complete library (Fig. S1). The complete set of supported capabilities is summarized in Table S1.

Rather than retaining the 79 specialized models as isolated generative checkpoints, Autocata converts them into reusable scientific capabilities through a domain-knowledge-driven registration process (Fig. 1b). In a model-centric regime, generative models exist as static checkpoints requiring manual script execution, configuration edits, and post-analysis data parsing. Autocata overcomes these limitations by embedding specialized models within machine-executable capability registries and automated workflow skills. Through a structured registry, each intermediate model is coupled with chemical rules, adsorbate validation, and explicit availability checks that govern its deployment within broader reaction networks. These models are dynamically integrated into reusable workflow skills, such as material-first multi-adsorbate screening, continuous sampling, native energy-window filtering, and automated report generation. By translating high-level scientific intents into executable computational pipelines, Autocata elevates standalone generative checkpoints into an interoperable, language-invocable scientific infrastructure for on-demand catalyst exploration. The agent autonomously coordinates adsorbate selection, pre-MLP material aggregation, and candidate evaluation without human intervention.

During online execution, Users interface with the system via desktop environments or the Feishu (https://www.feishu.cn/?from_site=lark) mobile interface (Fig. 1c). The orchestrating agent, powered by large language model (DeepSeek-V4), functions strictly as the planning and reasoning layer rather than the underlying computational engine, executing an automated eight-step operational pipeline. Upon receiving a natural-language prompt (Step 1), DeepSeek-V4 interprets task objectives and reaction constraints (Step 2), querying the scientific capability registry to select and compose matching domain models and workflow skills (Step 3). The selected capabilities are then deployed to execute adsorbate-conditioned structure generation coupled with geometric validity checks (Step 4), followed by MLP energy evaluations (Step 5). For multi-intermediate screening campaigns, Autocata aggregates shared materials across candidate reaction networks under consistent screening criteria (Step 6), continuously monitors execution progress (Step 7), and generates traceable reports with visual structural previews delivered directly back to the conversational interface (Step 8). This closed-loop orchestration enables Autocata to dynamically reconfigure execution paths and democratize expert-level catalyst discovery without task-by-task intervention from computational specialists.

## 2. Natural-language-guided single-adsorbate catalyst screening

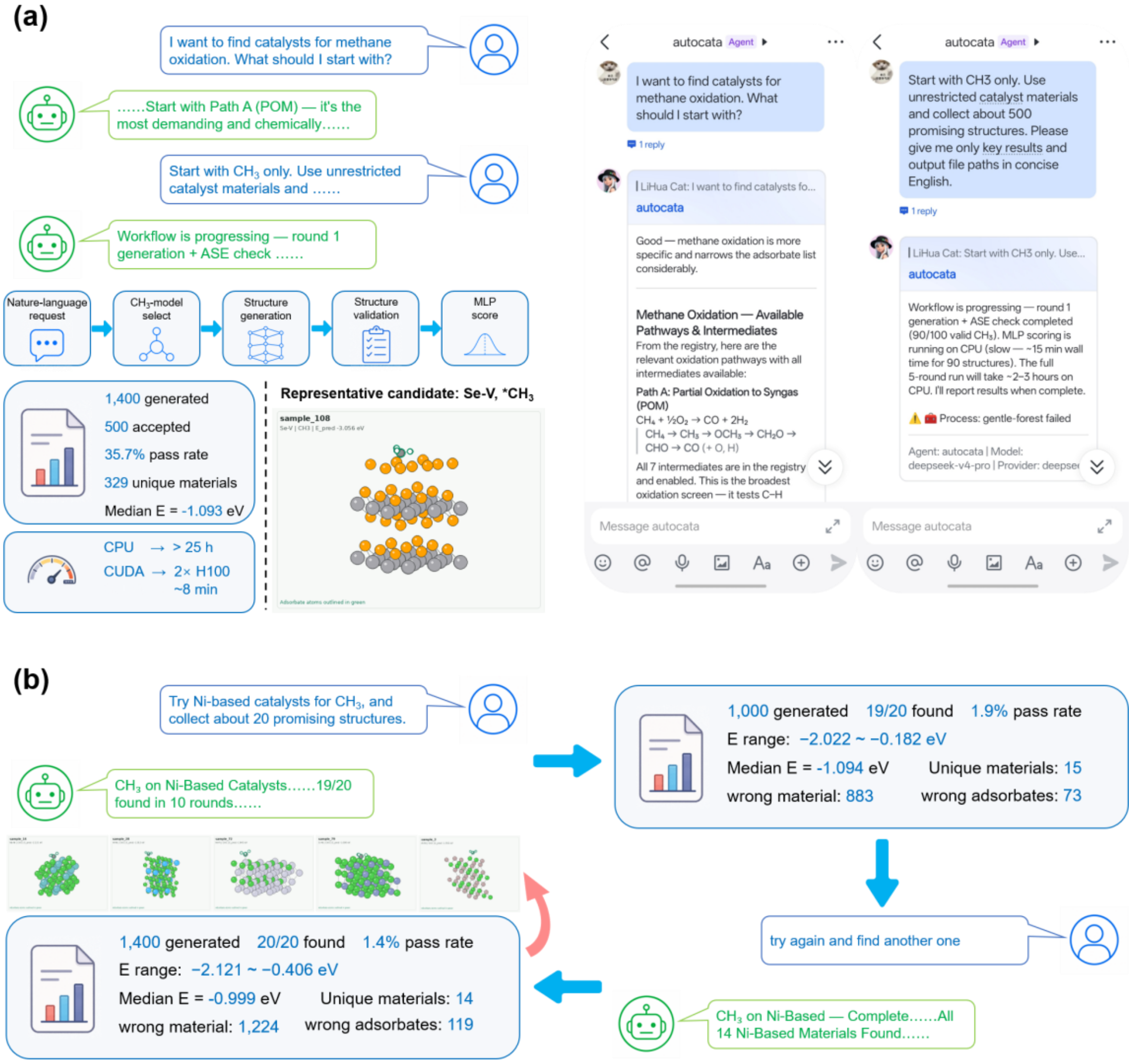

**Fig 2. Natural-language-guided $CH_3$ catalyst screening and constraint refinement.** (a) Open-ended *$CH_3$ screening initiated from a general methane-oxidation objective. Left: autonomous workflow schematic with CPU-to-GPU adaptation. Right: mobile-interface snapshots of natural-language interaction and real-time feedback. (b) Ni-based $CH_3$ screening under a user-defined compositional constraint, representative structures returned by the agent, indicated by the red arrow.

## Task 1: Objective-driven open-ended $CH_3$ catalyst screening

To evaluate whether Autocata could translate open-ended research intents into executable screening workflows, we first investigated methane oxidation as a representative single-adsorbate benchmark (Fig. 2). The campaign was initiated with a broad, conversational query: *"I want to find catalysts for methane oxidation. What should I start with?"* Rather than directly triggering a rigid, predefined computational script, the agent interpreted the high-level scientific objective and deconstructed possible reaction pathways, including partial oxidation to syngas[30], methane-to-methanol conversion[31], and oxidative coupling of methane[32] (Fig. 2a, Fig. S2, and Table S2). Guided by its encoded reaction knowledge and the available capabilities in the registry, Autocata identified *$CH_3$ as a tractable surface intermediate for initiating methane-activation screening. Upon receiving the user's refined instruction (*"Start with $CH_3$ only. Use unrestricted catalyst materials and collect about 500 promising structures."*), the agent autonomously configured and deployed the $CH_3$-specialized generative model, managing structure generation, geometric validity filtering, MLP scoring, and candidate ranking without manual script preparation.

Over seven sampling rounds, the automated campaign generated 1,400 candidate $CH_3$ adsorption configurations, yielding 500 validated candidates across 329 unique material compositions, corresponding to a 35.7% overall pass rate (Table S3). Predicted energies spanned from -3.056 eV to near zero, with a median value of -1.093 eV. The top 48 representative candidates ranked by MLP screening performance are detailed in Table S4. Promising leads included Se-V, Cd-S, and Cd-Cl, with Se-V exhibiting the most negative predicted *$CH_3$ adsorption energy (-3.056 eV). These rankings were used here to evaluate autonomous screening execution rather than to establish catalytic optimality. Crucially, this open-ended task highlighted Autocata's capability for dynamic execution management under operational bottlenecks. Identifying CPU-based MLP evaluation as a critical rate-limiting step with an estimated runtime exceeding 25 h, the agent accepted a natural-language user suggestion (*"I suggest CUDA if available when you calculate the MLP"*). Autocata interpreted this as an execution-level modification, reconfigured the MLP backend for GPU execution, and completed the evaluation in approximately 8 min using two NVIDIA H100 GPUs (Fig. S2).

**Task 2: Composition-constrained refinement of $CH_3$ catalyst screening**

To further assess whether Autocata could adaptively incorporate domain constraints into established workflows, we introduced a composition-specific refinement: *"Try Ni-based catalysts for $CH_3$, and collect about 20 promising structures."* Autocata interpreted this instruction as a material-composition filter and dynamically updated its candidate screening protocol without modifying the underlying generative model or requiring manual parameter re-configuration. During the initial search, 1,000 structures were generated over ten rounds, yielding 19 accepted Ni-containing $CH_3$ adsorption structures spanning 15 unique composition classes (Fig. 2b and Fig. S3). The majority of rejected candidates failed the newly enforced Ni compositional constraint (883 candidate failures), with a smaller subset failing adsorbate or MLP criteria. When prompted with a minimal follow-up instruction (*"Try again and find another one."*), Autocata avoided redundant workspace restarts and instead resumed execution from its stored state, expanding the search over 14 rounds and 1,400 generated candidates to successfully secure the target 20 Ni-based candidates. These final candidates exhibited predicted energies ranging from -2.121 eV to -0.406 eV (median -0.999 eV), featuring diverse formulations such as Nb-Ni, Hf-Ni, Ni-Pt, Cr-Ni, and Al-Ni (Fig. 2b and Tables S5-S6). Together, these single-adsorbate tasks demonstrate that Autocata integrates open-ended exploration with constraint-guided refinement, dynamically managing hardware allocation, state preservation, and workflow adaptation through concise natural-language feedback.

### 3. Natural-Language-Guided Reaction-Network Screening for Nitrogen Reduction

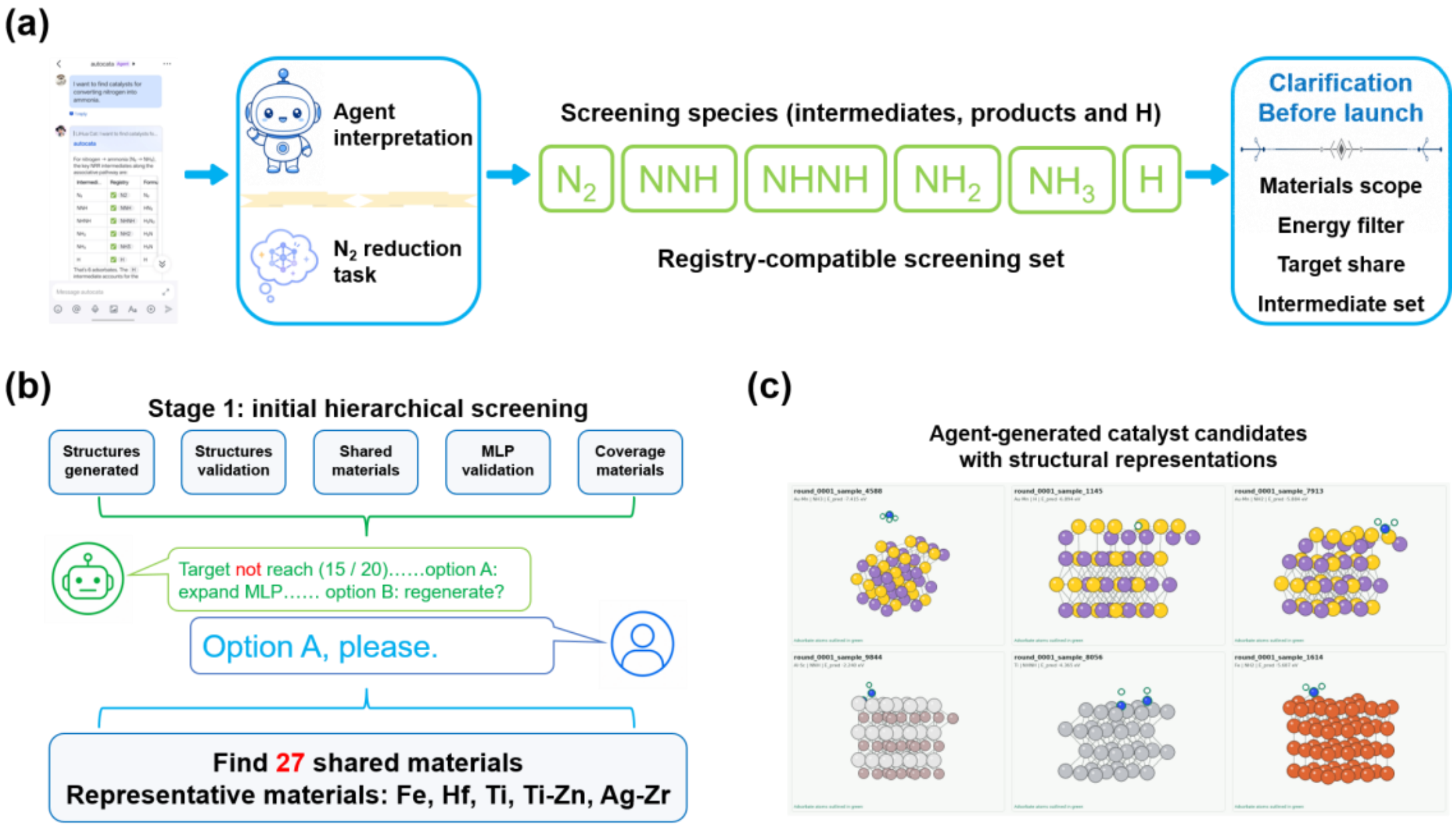

**Fig 3. Autonomous agent-driven reaction-network screening for $N_2$RR. (a)** Natural-language interpretation of an $N_2$RR discovery request. **(b)** Interactive hierarchical screening workflow involving structure generation, validation, shared-material identification, MLP-based energy evaluation, and adaptive strategy selection when predefined targets are not achieved. **(c)**

Representative catalyst structures generated and identified by the autonomous agent.

To evaluate whether Autocata could transform a broad reaction objective into an adaptive multi-intermediate screening campaign, we investigated the nitrogen reduction reaction ($N_2$RR) as a representative reaction-network benchmark (Fig. 3). Initiated by the expert-free request, *"I want to find catalysts for converting nitrogen into ammonia."*, the agent automatically mapped the objective to its capability registry without requiring predefined intermediates or explicit screening protocols. Autocata deconstructed the associative reduction pathway into five key nitrogen-containing intermediates ($N_2$, NNH, NHNH, $NH_2$, and $NH_3$) and additionally included adsorbed H to broaden the screening coverage of hydrogen-involving surface states relevant to the reduction environment (Fig. 3a and Fig. S4)[33,34]. Prior to execution, the agent converted the high-level intent into a structured campaign through a pre-launch clarification dialogue, confirming four critical operational parameters: catalyst material scope, energy filtration criteria, target number of shared materials, and intermediate species selection.

Upon user confirmation *"OK, launch,"* Autocata configured a material-first reaction-network screening strategy targeting 20 catalyst materials with MLP $< 0$ eV across all six adsorbates. During the initial sampling phase, approximately 60,000 structures were generated across the intermediate pool, with over 90% satisfying geometric and structural validity checks (Fig. S4). Rather than individually scoring all candidate structures, the agent executed a material-first aggregation step, identifying over 750 shared catalyst compositions across all six adsorbate pools. From this subset, an initial batch of 100 shared materials (comprising 14,198 structures) underwent MLP evaluation, yielding 15 materials that met the energy threshold across all six species (Table S7). Recognizing that the target of 20 shared materials was not achieved, Autocata analyzed the workflow status and localized the limitation to downstream MLP sampling depth rather than upstream structure generation. The agent dynamically presented two remediation pathways: Option A (expanding MLP evaluation on existing candidate materials) or Option B (regenerating new structural pools) (Fig. 3b).

After the user selected *"Option A, please,"* Autocata expanded the MLP evaluation set to 300 shared materials without repeating the computationally intensive generative sampling stage (Fig. 3b and Fig S4). Screening an additional 28,239 candidate structures identified 4,377 valid configurations, ultimately securing 27 catalyst materials with full coverage across all six adsorbates and surpassing the initial target of 20 materials (Fig. 3b and Tables S7-S8). Representative candidates included Fe, Hf, Ti, Ti-Zn, Ag-Zr, Au-Mn, Y, Au-Ti, V, and In-Ti. Beyond numerical filtering, the agent performed a qualitative chemical assessment on the resulting binding energies (Fig. S4). It flagged Fe and Au-Mn as strong-binding candidates prone to potential intermediate overbinding (Table S8), while highlighting Hf, Ti, Ti-Zn, Ag-Zr, In-Ti, and Cd-Y as systems exhibiting balanced adsorption interactions. This multi-intermediate campaign demonstrates that Autocata can coordinate multiple generative capabilities, diagnose workflow bottlenecks, and adaptively refine screening strategies through short natural-language interactions while providing meaningful domain interpretation.

## 4. Product-directed screening of the downstream $CO_2$-to-methanol pathway

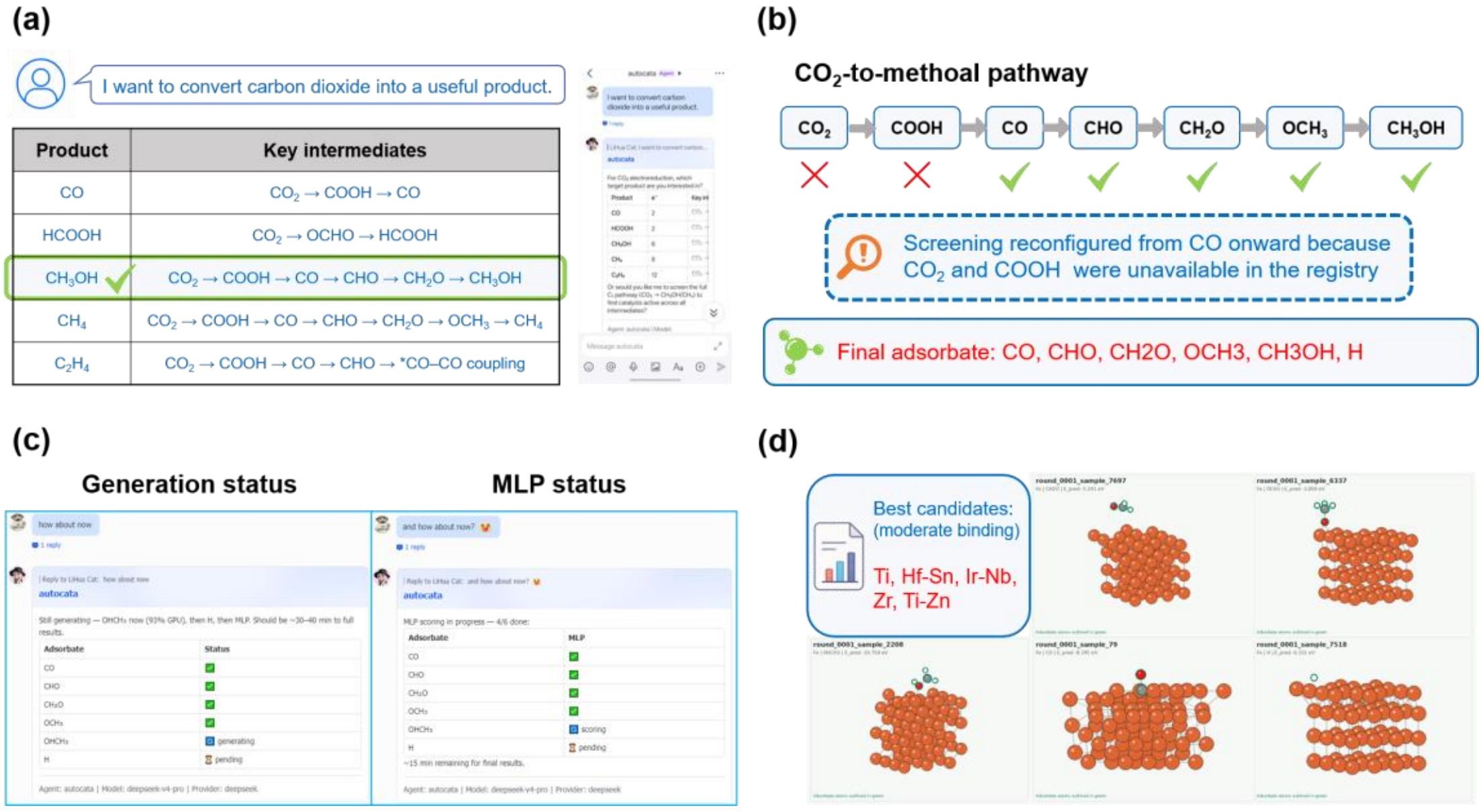

**Fig 4. Product-directed and registry-aware screening of a downstream $CO_2$-to-methanol pathway. (a)** Starting from a general $CO_2$-conversion query, the agent proposed several product-specific pathways, after which $CH_3OH$ was selected as the target product. **(b)** Mapping of the selected pathway to the available model registry. **(c)** Conversational monitoring of adsorbate-specific structure generation and MLP-scoring progress during execution. **(d)** Agent-highlighted candidates with moderate predicted binding and representative full-coverage adsorption structures returned by the workflow.

To assess whether Autocata could handle underspecified product objectives and adaptively manage model-registry limitations, we investigated the electrochemical $CO_2$ reduction reaction ($CO_2RR$)[35-38] as a complex multi-pathway benchmark (Fig. 4). The task was initiated with an open-ended request: *"I want to convert carbon dioxide into a useful product."* Recognizing that the target product was unspecified, the agent deconstructed the broad reaction space into five primary candidate endpoints (CO, HCOOH, $CH_3OH$, $CH_4$, and $C_2H_4$) along with their corresponding intermediate networks (Fig. 4a and Fig. S5). Upon receiving the follow-up instruction, "$CH_3OH$ sounds good. I want 20 shared catalyst materials.", Autocata queried its capability registry to configure the required intermediate models. Crucially, upon identifying that models for $CO_2$ and COOH were currently unavailable in the registry, the agent explicitly notified the user of this boundary and adaptively reconfigured the workflow into an executable downstream screening sequence beginning from adsorbed CO onward (Fig. 4b). The final screening set comprised six species: CO, CHO, $CH_2O$, $OCH_3$, $CH_3OH$, and adsorbed H as a hydrogen-related surface species.

Throughout execution, the conversational interface enabled transparent, real-time workflow monitoring without direct server access. In response to short status queries (*"how about now"*), the agent reported granular execution progress across individual generation tasks and downstream MLP evaluation stages (Fig. 4c and Fig. S5). The workflow generated 59,999 adsorption structures across the six selected adsorbates, with 53,445 passing geometric and physical validity filters. The valid fraction ranged from 70% for H to 98% for CO. Composition-level comparison identified 745 materials shared across all six candidate pools. Subsequent MLP scoring evaluated 28,072 candidates, with 7,463 structures satisfying the applied energy filter (Fig. S5) and yielding 91 full-coverage materials that far exceeded the user-defined target of 20 (Table S9).

Representative candidates and their corresponding surface adsorption geometries are shown in Fig. 4d, with the top 20 full-coverage materials listed in Table S10. Beyond raw numerical ranking, Autocata performed a qualitative chemical assessment, flagging Fe as a potential overbinding case due to its highly exothermic predicted adsorption energies, while highlighting systems with more moderate binding profiles, such as Ti, Hf-Sn, Ir-Nb, Zr, and Ti-Zn, as promising leads for subsequent validation (Fig. 4d and Fig. S5). This capability demonstrates that Autocata can resolve underspecified scientific intents, transparently navigate capability limits in the underlying model registry, reconfigure screening pathways, and deliver traceable candidates through intuitive conversational oversight.

## Conclusion

In summary, Autocata establishes an adaptive, capability-centric paradigm that transforms expert catalytic domain knowledge into a language-driven computational infrastructure. By uniting DeepSeek-V4 reasoning with a validated registry of 79 adsorbate-specialized generative capabilities, the agent autonomously executes end-to-end catalyst screening campaigns directly from high-level scientific prompts. The system's operational autonomy was demonstrated across three representative tasks, including compute-backend adaptation and state preservation in $CH_3$ screening, adaptive expansion of a six-species $N_2RR$ campaign, and capability-aware reformulation of downstream $CO_2$-to-methanol screening. While candidate catalysts remain subject to high-level electronic structure calculations and experimental synthesis, Autocata fundamentally transforms candidate discovery from a manual sequence of disconnected tools into an interactive, self-driving workflow. By shifting the core unit of scientific AI from isolated predictors to composable capabilities, this framework provides a universal architecture to serve as the cognitive orchestrator for fully autonomous, self-driving laboratories that bridge human intent with robotic synthesis. Ultimately, democratizing access to expert-level workflows and accelerating candidate discovery will empower researchers to rapidly address pressing global energy and environmental challenges.

## Methods

### Autonomous agent-enabled catalyst discovery workflow

Autocata was developed as an autonomous conversational agent that connects large-language-model-based task interpretation with computational modules for catalyst structure generation, validation, MLP-based evaluation, and result analysis. The agent converts natural-language research objectives into executable catalyst-screening workflows by interpreting user requests, identifying relevant adsorbates or reaction intermediates, selecting appropriate generative models, configuring screening parameters, and coordinating downstream computational procedures. The agent architecture consists of three functional layers: (i) a language-model-based planning layer responsible for task interpretation, reasoning, and interactive decision-making; (ii) a scientific computation layer containing adsorbate-specific generative models, structural validation modules, and MLP-based energy predictors; and (iii) an orchestration layer that manages communication between the language model and external computational tools. The orchestration framework was implemented using OpenClaw, which enables tool-based execution of scientific workflows through conversational interaction. The workflow is not restricted to a specific language model and can be configured with different API-accessible large language models supported by the orchestration framework. In all demonstrations presented in this work, DeepSeek-V4-Pro was used as the reasoning backend for task interpretation, workflow planning, model selection, and conversational interaction. The language model was not used for direct scientific prediction; instead, all structure generation and property evaluation were performed by dedicated computational models (Fig. S6).

**Generative model library construction**

Autocata integrates a library of adsorbate-specific catalyst structure generative models based on a Transformer architecture. Catalyst structures are represented as token sequences encoding lattice parameters, atomic species, and atomic coordinates, enabling the generative model to learn distributions of heterogeneous catalyst structures in a sequence-based representation. The generative model adopts a GPT-2 decoder-only architecture[12]. A general catalyst structure model was first pretrained using catalyst-adsorbate structures from the Open Catalyst 2020 (OC20) dataset[29]. The pretrained model was subsequently fine-tuned independently for different adsorbate species to generate adsorbate-specific catalyst structures. In total, 79 fine-tuned models were constructed, each corresponding to a specific adsorbate and capable of generating catalyst structures containing the target intermediate. During workflow execution, the agent selects the appropriate generative model according to the adsorbates or reaction intermediates identified from the user request. This model library enables Autocata to support both single-adsorbate catalyst discovery and multi-intermediate reaction-network screening. Detailed information regarding model architecture, pretraining procedure, fine-tuning datasets, and supported adsorbate species is provided in the Supporting Information.

**Adsorption-structure generation and sequence decoding**

Following adsorbate-model selection, the corresponding generative checkpoint and tokenizer were loaded by the structure-generation module. The implementation supports either a fully fine-tuned checkpoint or a parameter-efficient adapter loaded on top of the pretrained base model. Generation parameters, including the number of requested structures, maximum sequence length, top-k and nucleus-sampling thresholds, sampling temperature, batch size, and computing device, were read from the workflow configuration and could be overridden at execution time. An optional input prefix was tokenized and replicated across each generation batch. Candidate structures were then sampled autoregressively using stochastic decoding with top-k, top-p, and temperature controls. Generation continued until the configured maximum sequence length or the end-of-sequence token was reached. The generated sequences were truncated at the first end-of-sequence token, transferred to the CPU, and stored as serialized token-ID tensors for subsequent structure reconstruction. CUDA acceleration was selected automatically when available.

**Automated structure parsing and reconstruction**

The generated token sequences were decoded and reconstructed as periodic atomic structures using a two-stage procedure. First, the decoded sequence was passed to the structure-conversion utility to recover the simulation cell and atomic configuration and to obtain structural- and generation-validity flags. Second, element symbols and fractional coordinates were independently parsed and used to construct periodic ASE atoms objects[39]. Sequences that could not be converted into an atomic structure or did not contain valid element–coordinate tuples were recorded as parsing failures, whereas successfully reconstructed structures were exported in XYZ format. Each reconstructed structure was subsequently indexed according to its expected adsorbate formula. Under the sequence convention used here, the final $N$ atoms of each structure corresponded to the adsorbate, where $N$ was determined from the target molecular formula. Adsorbate-formula consistency was evaluated by comparing the elemental counts of these terminal atoms with the expected composition. The remaining atoms were assigned to the catalyst region. For candidate aggregation, a material label was defined as the alphabetically sorted set of unique elements present in the catalyst region. Thus, labels such as Fe and Hf-Sn represent elemental-composition classes rather than fixed stoichiometries, crystal phases, facets, or surface structures. Structure-level metadata, including the source XYZ file, adsorbate identity, adsorbate-composition status, material label, elemental set, and atom counts, were written to searchable CSV files. Material-level summary files recorded the total number of generated structures and adsorbate-valid structures associated with each elemental-composition label.

**Candidate organization and structure visualization**

Accepted structures were organized into candidate-level summary files containing the sample identifier, generation round, adsorbate identity, material label, structural file path, predicted energy, and available force statistics. The corresponding XYZ files were copied into a self-contained output directory and linked to a JSON manifest. To facilitate direct inspection of the returned candidates, Autocata generated both browser-based three-dimensional galleries and static structure previews. The interactive gallery displayed candidate metadata together with stick, sphere, and line representations and provided direct access to the associated XYZ files. For reaction-network tasks, representative previews could be restricted to materials satisfying the full-coverage criterion. Candidate structures were ordered by MLP results, with one representative initially selected from each material-adsorbate combination before additional structures were included. Static PNG images, rotating GIF files, and corresponding CSV and JSON manifests were generated for downstream reporting.

## Code availability

The codes using in this work are available at https://github.com/RileyLi911/autocata.

## Reference

1. Norskov, J. K. et al. The nature of the active site in heterogeneous metal catalysis. Chem. Soc. Rev. 37, 2163–2171 (2008).

2. Norskov, J. K., Bligaard, T., Rossmeisl, J. & Christensen, C. H. Towards the computational design of solid catalysts. Nat. Chem. 1, 37–46 (2009).

3. Calle-Vallejo, F. et al. Finding optimal surface sites on heterogeneous catalysts by counting nearest neighbors. Science 350, 185–189 (2015).

4. Chen, B. W. J., Xu, L. & Mavrikakis, M. Computational methods in heterogeneous catalysis. Chem. Rev. 121, 1007–1048 (2021).

5. Liu, L. & Corma, A. Metal catalysts for heterogeneous catalysis: From single atoms to nanoclusters and nanoparticles. Chem. Rev. 118, 4981–5079 (2018).

6. Motagamwala, A. H. & Dumesic, J. A. Microkinetic modeling: A tool for rational catalyst design. *Chem. Rev.* **121**, 1049–1076 (2021).

7. Greeley, J., Jaramillo, T. F., Bonde, J., Chorkendorff, I. & Nørskov, J. K. Computational high-throughput screening of electrocatalytic materials for hydrogen evolution. *Nat. Mater.* **5**, 909–913 (2006).

8. Ulissi, Z. W., Medford, A. J., Bligaard, T. & Nørskov, J. K. To address surface reaction network complexity using scaling relations machine learning and DFT calculations. Nat. Commun. 8, 14621 (2017).

9. Zhao, Q., Xu, Y., Greeley, J. & Savoie, B. M. Deep reaction network exploration at

a heterogeneous catalytic interface. Nat. Commun. 13, 4860 (2022).

10. Unke, O. T. et al. Machine learning force fields. Chem. Rev. 121, 10142–10186 (2021).

11. Tran, K. & Ulissi, Z. W. Active learning across intermetallics to guide discovery of electrocatalysts for $CO_2$ reduction and $H_2$ evolution. Nat. Catal. 1, 696–703 (2018).

12. Mok, D. H. & Back, S. Generative pretrained transformer for heterogeneous catalysts. J. Am. Chem. Soc. 146, 33712–33722 (2024).

13. Vandermause, J., Xie, Y., Lim, J. S., Owen, C. J. & Kozinsky, B. Active learning of reactive bayesian force fields applied to heterogeneous catalysis dynamics of H/pt. Nat. Commun. 13, 5183 (2022).

14. Ghanekar, P. G., Deshpande, S. & Greeley, J. Adsorbate chemical environment-based machine learning framework for heterogeneous catalysis. Nat. Commun. 13, 5788 (2022).

15. Pablo-García, S. et al. Fast evaluation of the adsorption energy of organic molecules on metals via graph neural networks. Nat. Comput. Sci. 3, 433–442 (2023).

16. Zeni, C. et al. A generative model for inorganic materials design. Nature 639, 624–632 (2025).

17. Song, Z. et al. Inverse design of promising electrocatalysts for CO2 reduction via generative models and bird swarm algorithm. Nat. Commun. 16, 1053 (2025).

18. Antunes, L. M., Butler, K. T. & Grau-Crespo, R. Crystal structure generation with autoregressive large language modeling. Nat. Commun. 15, 10570 (2024).

19. Li, R. et al. Generative intelligence explores the chemical space of ten million

catalysts. Chem. Sci. 17, 12996–13006 (2026).

20. Li, R. et al. Reaction-network-level discovery of ammonia synthesis catalysts via ten-million-scale generative exploration, arXiv:2606.22926 (2026).

21. Shen, Y. et al. HuggingGPT: Solving AI tasks with ChatGPT and its friends in hugging face. in ADVANCES IN NEURAL INFORMATION PROCESSING SYSTEMS 36 (NEURIPS 2023) (eds Oh, A. et al.) (Neural Information Processing Systems (nips), New Orleans, LA, 2023).

22. Gao, L. et al. PAL: Program-aided language models. in INTERNATIONAL CONFERENCE ON MACHINE LEARNING, VOL 202 (eds Krause, A. et al.) vol. 202 (Jmlr-Journal Machine Learning Research, Honolulu, HI, 2023).

23. Yao, S. et al. ReAct: Synergizing reasoning and acting in language models. Preprint at https://doi.org/10.48550/arXiv.2210.03629 (2023).

24. Schick, T. et al. Toolformer: Language models can teach themselves to use tools. in ADVANCES IN NEURAL INFORMATION PROCESSING SYSTEMS 36 (NEURIPS 2023) (eds Oh, A. et al.) (Neural Information Processing Systems (nips), New Orleans, LA, 2023).

25. M. Bran, A. et al. Augmenting large language models with chemistry tools. Nat. Mach. Intell. 6, 525–535 (2024).

26. Boiko, D. A., MacKnight, R., Kline, B. & Gomes, G. Autonomous chemical research with large language models. Nature 624, 570–578 (2023).

27. Ock, J., Meda, R. S., Vinchurkar, T., Jadhav, Y. & Barati Farimani, A. Adsorb-Agent: autonomous identification of stable adsorption configurations via a large language model agent. Digital Discovery 5, 617–629 (2026).

28. Rothfarb, S. et al. Hierarchical multi-agent large language model reasoning for autonomous heterogeneous catalyst discovery. npj Comput. Mater. 12, 309 (2026).

29. Chanussot, L. et al. Open Catalyst 2020 (OC20) Dataset and Community Challenges. ACS Catal. 11, 6059–6072 (2021).

30. Enger, B. C.; Lødeng, R.; Holmen, A. A review of catalytic partial oxidation of methane to synthesis gas with emphasis on reaction mechanisms over transition metal catalysts. Appl. Catal. A 346, 1–27 (2008).

31. Ravi, M.; Ranocchiari, M.; van Bokhoven, J. A. The Direct Catalytic Oxidation of Methane to Methanol—A Critical Assessment. Angew. Chem. Int. Ed. 56, 16464–16483 (2017).

32. Schwach, P.; Pan, X.; Bao, X. Direct Conversion of Methane to Value-Added Chemicals over Heterogeneous Catalysts: Challenges and Prospects. Chem. Rev. 2017, 117, 8497–8520 (2017).

33. Skúlason, E. et al. A theoretical evaluation of possible transition metal electro-catalysts for $N_2$ reduction. Phys Chem Chem Phys 14, 1235–1245 (2012).

34. Montoya, J. H.; Tsai, C.; Vojvodic, A.; Nørskov, J. K. The Challenge of Electrochemical Ammonia Synthesis: A New Perspective on the Role of Nitrogen Scaling Relations. ChemSusChem 8, 2180–2186 (2015).

35. Peterson, A. A., Abild-Pedersen, F., Studt, F., Rossmeisl, J. & Nørskov, J. K. How copper catalyzes the electroreduction of carbon dioxide into hydrocarbon fuels. Energy Environ. Sci. 3, 1311 (2010).

36. Nitopi, S. et al. Progress and perspectives of electrochemical $CO_2$ reduction on copper in aqueous electrolyte. Chem. Rev. 119, 7610–7672 (2019).

37. Hitt, J. L. et al. A high throughput optical method for studying compositional effects in electrocatalysts for $CO_2$ reduction. Nat. Commun. 12, 1114 (2021).

38. Shi, G. et al. Nanoconfinement promotes $CO_2$ electroreduction to methanol on a molecular catalyst. Nat. Commun. 16, 7359 (2025).

39. Hjorth Larsen, A. *et al.* The atomic simulation environment—a python library for working with atoms. *J. Phys. Condens. Matter* **29**, 273002 (2017).

## Acknowledgements

This work is supported by National Natural Science Foundation of China (92477105, 92577120), Shanghai Municipal Science and Technology Major Project, and Foundation of the Key Laboratory of Low-Carbon Conversion Science & Engineering, Shanghai Advanced Research Institute, Chinese Academy of Sciences (KLLCCSE-202201Z, SARI, CAS). R. Q. thanks for the Innovation Program of Shanghai Advanced Research Institute, CAS (2025CP007). All calculations were performed at National Supercomputing Center in Shanghai.

## Author contributions

R.Q. initiated the project. B.Z. and Y.G. supervised the project. R.L. performed the calculations and data analysis. All authors participated in the discussions.

## Competing interests

The authors declare no competing interests.